# A Second-Order Symplectic Integrator for Guiding-Center Equations
JR Cary, U. Colorado
February 7, 1989

## I. Introduction

The purpose of this note is to describe the elements of a symplectic integrator. This integrator is designed to preserve the symplectic form (i.e., Poincare's integral invariants) to numerical accuracy. It is second order in the time step: the error after integrating for a finite time is $O(\Delta t^2)$. To begin, we recall Hamilton's equations in Sec. II. The integrator, which requires a nonlinear equation solve at each step, is discussed in Sec. III. A fast method for performing this solve is described in Sec. IV. Notes on the submission of this paper to Arxiv are given in the appendix.

This integrator is being developed for integration of canonical guiding-center equations of motion. The guiding-center equations are canonical only in flux coordinates of integrable magnetic fields. Therefore, Sec. V contains a discussion of the coordinates for such fields. Finally, Sec. VI discusses the possible sets of guiding- center equations.

## II. Hamilton's equations and canonical variables

Hamilton's equations describe the motion of canonical variables **(q,p).** Here, we will assume that the phase space has n degrees of freedom, so that the vectors **q** and **p** each have n components. A phase space point is denoted by the vector z, which is 2n-dimensional. The components of z are those of **q** followed by those of **p,** i.e.

$$z_i = \begin{cases} q_1 \text{ for } 1 \leq i \leq n \\ p_i \text{ for } n+1 \leq i \leq 2n \end{cases} \quad (1)$$

The evolution of these coordinates is determined by the Hamiltonian, **H(q,p),** via Hamilton's equations:

$$\dot{q}_i = \frac{\partial H}{\partial p_i} \quad (2a)$$

and

$$\dot{p}_i = -\frac{\partial H}{\partial q_i} \quad (2b)$$

Hamilton's equations have several nice properties: they conserve the Hamiltonian if it is time-independent (as is assumed here), they preserve the integral invariants of Poincare, and they preserve the momenta corresponding to ignorable coordinates.

An alternate way of writing Hamilton's equations is

$$\dot{z}_i = \sum_{j=1}^{2n} J_{ij} \frac{\partial H}{\partial z_j} \quad (3)$$

in which J, the Poisson tensor, is given by

$$J_{ij} = \delta_{i,j-n} - \delta_{i-n,j}. \quad (4)$$

That is, in nxn blocks, J has the form

$$\mathbf{J} = \begin{pmatrix} \mathbf{0} & \mathbf{I_n} \\ -\mathbf{I_n} & \mathbf{0} \end{pmatrix}. \tag{5}$$

with $\mathbf{I_n}$ being the nxn identity matrix. Because **J** has this simple form, matrix multiplies involving **J** are essentially just interchanges, and, hence, do not require the computational effort of a usual matrix multiply.

## III. A symplectic algorithm for Hamilton's equations

Hamilton's equations generate a *symplectic* transformation, one that preserves Poincare's integral invariants. This means that the Poisson Brackets of the time- advanced variables evaluated with respect to the initial conditions obey the canonical relations. However, in general, finite time-step approximations to Hamilton's equations, such as Euler's method,

$$\mathbf{z}(t + \Delta t) = \mathbf{z}(t) + \Delta t \mathbf{J} \cdot \frac{\partial H}{\partial \mathbf{z}}. \tag{6}$$

do not yield symplectic transformations. Symplectic transformations are desired, because they preserve the long-time structure of phase space even for reason'ably large time step. In contrast, nonsymplectic algorithms generally cause slow diffusion of rigorously conserved invariants [1].

There are now many symplectic integrators available. Recently, I have discovered a new symplectic integrator that is second-order accurate. I will not go into the details of the proof. I will simply state the result. The difference vector, $\zeta$, is defined by

$$\boldsymbol{\zeta} \equiv \mathbf{z}(t + \Delta t) - \mathbf{z}(t). \tag{7}$$

The difference vector is found by solving the nonlinear ~~differential~~ equation,

$$\mathbf{F}(\zeta) \equiv \zeta - \Delta t \mathbf{J} \cdot \frac{\partial H}{\partial \mathbf{z}}\left(\mathbf{z} + \tfrac{1}{2}\zeta\right) = 0. \tag{8}$$

Upon solving Eq. (8) for the difference vector, Eq. (7) is used to calculate the value of the new point, $\mathbf{z}(t+\Delta t)$.

The drawback of this method is that the difference vector is defined implicitly. Thus, it must be found iteratively via Newton's method or straightforward iteration. In the next section, we describe a fast and simple method for obtaining such a solution.

## IV. Algorithm for finding a solution to a set of nonlinear equations

Newton's method is commonly used for finding a solution to a set of nonlinear equations, $F_j(\zeta=(\zeta_1,...\zeta_{2n}))=0$. Newton's method has the property that the error squares at each iteration. If the error prior to the iteration is $O(\varepsilon)$, then after the iteration the error is $O(\varepsilon^2)$. This is known as superconvergence. This property leads to rapid convergence to the answer. For example, if the error initially is $10^{-2}$, then convergence to CRAY accuracy ($10^{-14}$) is obtained after only three iterations. The problem with Newton's method is that it requires the knowledge of the (Jacobian) derivative matrix,

$$M_{ij} = \frac{\partial F_i}{\partial \zeta_j}. \tag{9}$$

This implies that 4n2 quantities must be calculated at each iteration

Recently, improvements to Newton's method have been found. The basis of these improvements is that one can obtain (almost) superconvergence without completely recalculating the Jacobian. (Almost means that the error mapping is $\varepsilon \rightarrow \varepsilon^{1.6}$.) Instead one need only update the Jacobian via Broyden's method [2,3]. Moreover, the updates involve only the data known from previous iterations. I will now describe an algorithm for solving equations using this method.

### A. Initialization

We begin by assuming we have an approximate derivative matrix M. This matrix is either (1) exactly calculated, (2) the value of M from the last integration step of the differential equation solver, or (3) simply set to the identity. Switches should be put into the coding to allow different ways of initializing M both at the first time step and at subsequent time steps.

In addition, we assume that we have some initial guess for $\zeta$, which is the difference vector corresponding to this step. Because $\zeta$ varies little from step to step, a good initial guess is the (converged) value from the last integration step. Alternatively, one may choose $\zeta = 0$. Again, switches should be implemented in the code for testing. Lastly, one must have the first value of the residual, $R=F(\zeta)$.

### B. Linear system solve for the change of the difference vector.

According to Newton's method, the change s of the difference vector is found from solving the linear system of equations,

$$\sum_{j=1}^{2n} M_{ij} s_j = -R_i. \tag{10}$$

From s, one finds the new estimate of the difference vector,

$$\zeta_i = \zeta_i + s_i. \tag{11}$$

At this point, one must check the magnitude of s. If it is small, i.e. less than machine accuracy, the integration step is finished. Otherwise, one must proceed to *C* below.

### C. Update

At this point, the residual vector and the derivative matrix must be updated. The new residual is given by,

$$R = F(\zeta). \tag{12}$$

To calculate the updated derivative matrix, one must first calculate the normalization factor for the change,

$$\alpha \equiv \left(\sum_{i=1}^{2n} s_i^2\right)^{-1}. \tag{13}$$

From this one obtains the derivative matrix,

$$M_{ij} = M_{ij} + \alpha R_i s_j. \tag{14}$$

This completes the iteration. One must now return to step B.

## V. Magnetic coordinates

The symplectic integrator is being developed for the purpose of integrating the guiding center equations of motion. As so far developed, the symplectic integrator works only for canonical coordinates. Therefore, we are unable to integrate the general equations of Littlejohn [4]. Instead we must keep within the framework of integrable magnetic fields (those with nested flux surfaces), for which canonical coordinates are known to exist [5,6]. Integrable magnetic fields are most conveniently described in terms of flux coordinates. We describe those in this section. We follow the conventions of Cary et al [7].

The components of the vector potential in the covariant representation,

$$\mathbf{A} = \psi \nabla \theta + A_\varphi(\psi) \nabla \varphi. \tag{15}$$

are flux surface quantities, i.e. functions of only the toroidal flux, $\psi$. This vector potential implies that the Clebsch (2-form) representation of the magnetic field is

$$\mathbf{B} = \nabla \psi \times \nabla \theta + \frac{dA_\varphi}{d\psi} \nabla \psi \times \nabla \varphi. \tag{16}$$

From this representation we see that the rotational transform, $\iota = d\theta/d\varphi$, is given by

$$\iota = -\frac{dA_\varphi}{d\psi}. \tag{17}$$

For integration of the equations of motion, we will need an explicit form for the rotational transform and the poloidal flux, $A\varphi$. A reasonable tokamak profile for the rotational transform is

$$\iota = \iota_0 + \iota_1 (\psi/\psi_e). \tag{18a}$$

where $\psi_e$ is the value of the toroidal flux at the edge. Typically, $\iota_0 = 1$, and $\iota_1 = -3/4$ for $q=1/\iota$ to be 4 at the edge. Integration of Eq. (17) then gives

$$A_\varphi = -\iota_0 \psi - \frac{1}{2} \iota_1 (\psi^2/\psi_e). \tag{18b}$$

To attribute some physical meaning to these variables, we introduce $B_0$. the value of the magnetic field on axis. Then we can define a flux-surface edge-radius,

$$a = (2\psi_e/\bar{B}_0)^{1/2}, \tag{19}$$

which should have a value close to that of the mean edge radius. Similarly we define a flux-surface averaged minor radius,

$$r = (2\psi/\bar{B}_0)^{1/2}, \tag{20}$$

and a normalized radius,

$$\rho = r/a, \tag{21}$$

which varies between zero and unity in the plasma.

The guiding-center Hamiltonian also involves the covariant components of B. In the Boozer representation, the angular components of B are flux-surface quantities,

$$\mathbf{B} = B_\psi(\psi,\theta,\varphi)\nabla\psi + B_\theta(\psi)\nabla\theta + B_\varphi(\psi)\nabla\varphi, \tag{22}$$

Again, we must have explicit forms for these functions. For a crude model, it is sufficient to ignore $B_\psi$. Let us define $R_0$ to be the major radius. Then, to lowest order in $\beta$ one can show that

$$B_\varphi = R_0 B_0, \tag{23}$$

a constant. Finally we need $B_\theta$. The rotational transform is given by $\iota = \mathbf{B}\cdot\nabla\theta/\mathbf{B}\cdot\nabla\varphi$. With the approximations that (1) the flux coordinates are nearly orthogonal, and (2) the surfaces are nearly circular, we obtain

$$B_\theta = \frac{r^2}{R_0^2}\iota B_\varphi. \tag{24}$$

The last quantity we need for the guiding-center equations of motion is the magnetic field strength. For the large-aspect-ratio, nearly circular case, it is appropriate to use the form,

$$B = B_0[1 - \varepsilon_t \cos(\theta) - \varepsilon_r \cos(N\varphi)]. \tag{25}$$

The toroidicity factor is, as usual, given by the lowest-order term of the aspect ratio expansion,

$$\varepsilon_t = r/R_0. \tag{26}$$

I have yet to work out the ripple term. We need to investigate what others have done.

## VI. Guiding-center equations of motion

We must now choose a set of variables for the integration. There are at least two possibilities. Reference 5 contains what I call the $\theta_0$-formalism, so called because one of the variables is the value of the poloidal angle at the point where the field line of the guiding center intersects $\varphi=0$. Reference 6 contains the $p_\theta$-$p_\varphi$ formalism, named for the variables of this formalism. Both sets of coordinates have advantages and drawbacks, so we should discuss them further. I prefer the formalism because the variables have the correct periodicity. However, the $\theta_0$-formalism may be simpler computationally.

A noncanonical set of guiding-center coordinates are the flux variables ($\psi,\theta,\varphi$) combined with the parallel velocity u. The equations of motion for these variables may be obtained by variation of the phase-space Lagrangian,

$$L = mu\left(\frac{B_\psi}{B}\right)\dot{\psi} + \left(\frac{e\psi}{c} + \frac{muB_\theta}{B}\right)\dot{\theta} + \left(\frac{eA_\varphi}{c} + \frac{muB_\varphi}{B}\right)\dot{\varphi} - H. \tag{27}$$

for which the Hamiltonian is

$$H = \tfrac{1}{2}mu^2 + V. \tag{28}$$

in terms of the guiding-center potential,

$$V = \mu B + e\Phi. \tag{29}$$

The equations of motion derived from variation of the phase-space Lagrangian (27) are not of Hamilton's form because the coordinates are noncanonical. However, when $B_\psi$ vanishes, the momenta,

$$p_\theta = e\psi/c + muB_\theta/B. \tag{30a}$$

and

$$p_\varphi = eA_\varphi/c + muB_\varphi/B. \tag{30b}$$

are canonically conjugate to the angles, $(\theta,\varphi)$. Thus, the full set of variables is $(\theta,\varphi,p_\theta,p_\varphi)$

To calculate the Hamiltonian and its derivatives from $p_\theta$ and $p_\varphi$, one must solve Eqs. (30) for the parallel velocity and the flux $\psi$. Given the flux, the parallel velocity may be easily calculated from either of Eqs. (30), since they are both linear in the parallel velocity. However, to find the flux, one must solve the equation,

$$p_\theta B_\varphi - p_\varphi B_\theta = (e/c)(\psi B_\varphi - A_\varphi B_\theta). \tag{31}$$

obtained from Eqs. (30) by elimination of u. For the general case this will again require Newton iteration. However, for our model this equation can be solved explicitly since it is quartic in the flux. It can be further simplified by dropping the last term, which is smaller by $(r/R_0)^2$. This yields an equation that is quadratic in $\psi$, but we must understand the implications of this approximation.

Alternatively, one may use the $\theta_0$-formalism. In this formalism, the coordinates are $(\theta_c,\chi)$ with canonically conjugate momenta $(e\psi/c,\rho)$ respectively, which are defined by

$$\theta_c = \theta - \iota\varphi - \rho\left(B_\psi - \theta\frac{\partial B_\theta}{\partial \psi} - \varphi\frac{\partial B_\varphi}{\partial \psi}\right), \tag{32a}$$

$$\rho = muc/eB, \tag{32b}$$

and

$$\chi = e(\theta B_\theta + \varphi B_\varphi)/c. \tag{32c}$$

To evaluate the Hamiltonian and its derivatives, one must solve Eqs. (32a) and (32c) for $\theta$ and $\varphi$ as functions of $\chi$ and $\theta_c$. If we neglect $B\psi$ as before, then these equations are linear, and so they may be solved explicitly. Hence, it would seem that these equations are more amenable to symplectic integration. However, I dislike the fact that the variables $\theta_c$ and $\chi$ do not have the periodicity of the problem, i.e. they do not increase by a multiple of $2\pi$ when $\theta$ and $\varphi$ increase by a multiple of $2\pi$.

## V. Bibliography

1. See, e.g., PJ Channel and C Scovel, *Symplectic Integration of Hamiltonian Systems*, Los Alamos preprint LA-UR-88-1828, submitted to Nonlinearity (1988).
2. Broyden, Charles G. "A class of methods for solving nonlinear simultaneous equations." Mathematics of computation 19.92 (1965): 577-593.
3. Dennis, J. E. "On the convergence of Broyden's method for nonlinear systems of equations." Mathematics of Computation 25.115 (1971): 559-567.
4. Littlejohn, Robert G. "Variational principles of guiding centre motion." Journal of Plasma Physics 29.1 (1983): 111-125.

5. White, R. B., Allen H. Boozer, and Ralph Hay. Drift Hamiltonian in magnetic coordinates. No. PPPL--1880. Princeton Univ., 1982.
6. White, R. B., and M. S. Chance. "Hamiltonian guiding center drift orbit calculation for plasmas of arbitrary cross section." The Physics of fluids 27.10 (1984): 2455-2467.
7. Cary, John R., C. L. Hedrick, and J. S. Tolliver. "Orbits in asymmetric toroidal magnetic fields." The Physics of fluids 31.6 (1988): 1586-1600.

## Appendix: History of this paper

This paper was originally sent to Jim Rome of Oak Ridge National Laboratory in 1989 as part of a collaboration with him on guiding center modeling for transport. Interest in this method resurfaced, and so it was decided in September, 2018 to put this on Arxiv so that it could be cited. The electronic version was since lost, so the paper copy was transcribed verbatim into this electronic form, which is now posted at arxiv.org.